\documentclass[utf8]{frontiersSCNS} 

\usepackage{url,hyperref,microtype,subcaption}
\usepackage[onehalfspacing]{setspace}

\def\keyFont{\fontsize{8}{11}\helveticabold }
\def\firstAuthorLast{Sesana} 
\def\Authors{Alberto Sesana\,$^{1,*}$}
\def\Address{$^{1}$ Department of Physics G. Occhialini, University of Milano - Bicocca, Piazza della Scienza 3, 20126 Milano, Italy}

\def\corrAuthor{Corresponding Author}
\def\corrEmail{alberto.sesana@unimib.it}

\newcommand{\msun}{{\rm M}_{\odot}}

\begin{document}
\onecolumn
\firstpage{1}

\title[Black hole science with LISA]{Black hole science with the Laser Interferometer Space Antenna} 

\author[\firstAuthorLast ]{\Authors} 
\address{} 
\correspondance{} 

\extraAuth{}

\maketitle

\begin{abstract}

I review the scientific potential of the Laser Interferometer Space Antenna (LISA), a space-borne gravitational wave (GW) observatory to be launched in the early 30s'. Thanks to its sensitivity in the milli-Hz frequency range, LISA will reveal a variety of GW sources across the Universe, from our Solar neighbourhood potentially all the way back to the Big Bang, promising to be a game changer in our understanding of astrophysics, cosmology and fundamental physics. This review dives in the LISA Universe, with a specific focus on black hole science, including the formation and evolution of massive black holes in galaxy centres, the dynamics of dense nuclei and formation of extreme mass ratio inspirals, and the astrophysics of stellar-origin black hole binaries. 
  
\tiny
\keyFont{ \section{Keywords:} gravitational waves, black hole physics, binary systems, cosmology, tests of gravity} 

\end{abstract}

\section{Introduction}

\begin{figure}
\centering
\includegraphics[width=6.5in,angle=0.]{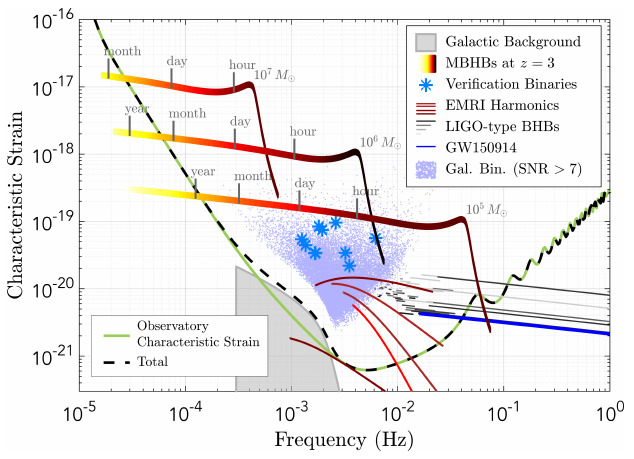}
\caption{The dimensionless 'characteristic strain' of GW sources in the LISA frequency band. The nominal detector sensitivity is shown by the green line. Displayed are tracks of three equal MBHBs at $z=3$ with total masses of $10^5, 10^6, 10^7\msun$, the first five harmonics of an EMRI at $z=1.2$ (red solid lines), a sample of stellar-mass BHBs (black solid lines) and several thousands of resolvable galactic binaries (blue dots). The subset of known 'verification binaries' is shown with blue asterisks. The 'confusion noise' arising from the millions of galactic binaries that cannot be resolved individually is shown by the grey shaded area. \citep[From][]{2017arXiv170200786A}.}
\label{fig:lisa_sources}
\end{figure}

Despite the wealth of revolutionary results already delivered \citep{2019PhRvX...9c1040A}, gravitational wave (GW) astronomy is still in its infancy. LIGO \citep{2009RPPh...72g6901A} and Virgo \citep{2015CQGra..32b4001A} are in fact only sensitive to binary systems of $\lesssim 100\msun$ out to $z\approx 1$, leaving us still blind to the vast majority of GW sources in the Universe. This will profoundly change within the next two decades, when GW revelation instruments and techniques will access sources covering a much larger spectrum of masses (up to $10^{10}\msun$) essentially anywhere in the Universe. The 3G detectors Einstein Telescope \citep{2010CQGra..27s4002P} and Cosmic Explorer \citep{2019BAAS...51g..35R} will cover the Hz to kilo-Hz frequency range, populated by binaries of compact objects (CO) of different nature, out to high redshift. Neutron star binaries (NSBs) will be observed out to $z>2$ at a rate of tens of thousands per year, and similar rates are expected for black hole binaries (BHBs) which will be observable out to $z\approx 20$ \citep{2014JPhCS.484a2008V}. Interestingly, the extension of the sensitivity window down to few Hz, will open-up the uncharted land of intermediate mass black holes \citep{2020NatAs...4..260J}. At the opposite end of the frequency and source  mass spectrum, radio millisecond pulsar data, collected and analysed by pulsar timing array \citep[PTA,][]{1990ApJ...361..300F} collaborations \citep{2016MNRAS.458.3341D,2018ApJS..235...37A,2020PASA...37...20K}, are the gateway to the $\mu-$Hz to nano-Hz frequency range. Here, the expectation is to detect a stochastic GW background (GWB) emerging from the incoherent superposition of signals from a cosmic population of massive black hole binaries (MBHBs), forming in the aftermath of galaxy mergers occurring along the assembly of cosmic structures \citep{2008MNRAS.390..192S,2012ApJ...761...84R}. The international PTA \citep[IPTA,][]{2016MNRAS.458.1267V} is working in this direction and with the advent of the Square Kilometre Array \citep[SKA,][]{2009IEEEP..97.1482D}, there is also the expectation to resolve the most massive inspiralling individual MBHBs in the universe \citep{2009MNRAS.394.2255S,2018MNRAS.477..964K}. 

The bridging milli-Hz frequency window will be explored from space, thanks to the Laser Interferometer Space Antenna \citep[LISA][]{2017arXiv170200786A}, one of the next large missions of the European Space Agency with the participation of NASA, to be flown in the early 30s'. Being sensitive to the milli-Hz frequency band, from $\approx 0.1\,$milli-Hz to 0.1$\,$Hz, LISA  is ideally suited to probe GW sources across the mass and distance scales, from the solar neighbourhood to the Big Bang. Starting from our backyard, contrary to ground based detectors and PTAs, LISA is expected to observe a bonanza of sources within the Milky Way (MW). Those include millions of galactic compact objects (COs), mostly double white dwarfs (DWDs), building up an unresolved confusion noise around 0.5-2 milli-Hz \citep{2001A&A...375..890N}. Up to 20k such DWDs will be individually resolvable \citep{2012ApJ...758..131N}, along with several tens of NSBs \citep{2020MNRAS.492.3061L} and few BHBs \citep{2016MNRAS.460L...1S,2020MNRAS.494L..75S}. Moreover, LISA has the unique potential to detect the presence of planets around nearby DWDs \citep{2019NatAs...3..858T} and perhaps dozens of brown dwarfs and sub-stellar objects orbiting SgrA$^{*}$ \citep{2003ApJ...583L..21F}, known as X-MRI \citep{2019PhRvD..99l3025A}. LISA will detect many more BHBs outside the MW, being sensitive to the early inspiral of these systems centuries to weeks before they enter the ground based detector sensitivity band, out to $z\approx 0.5$ \cite{2016PhRvL.116w1102S}. COs inpiralling onto MBHs, known as extreme mass ratio inspirals (EMRIs) can be detected out to $z\approx 2$ \cite{2017PhRvD..95j3012B}, whereas coalescing massive black hole binaries MBHBs in the mass range $10^4\msun<M<10^7\msun$ can be seen anywhere in the Universe \cite{2016PhRvD..93b4003K}. Last but not least, the frequency range covered by LISA makes it sensitive to TeV energy scales, where a stochastic GWB might be produced in the early Universe by, e.g., first order phase transitions or cosmic defects like strings and loops. A visual summary of selected LISA sources is depicted in figure \ref{fig:lisa_sources}, from \cite{2017arXiv170200786A}. The observation of each class of sources will provide invaluable insights in astrophysics, cosmology and fundamental physics, which is beyond what can be reasonably tackled within the few pages of this review. We therefore focus on a subset of sources, specifically MBHBs, EMRIs and BHBs, highlighting their astrophysical potential in particular. The payouts of studying fundamental physics with low frequency GWs are extensively described in a dedicated LRR article \cite{2013LRR....16....7G}, whereas a comprehensive review of cosmological GWBs with much focus on LISA can be found in \citep{2018CQGra..35p3001C}.

\section{Massive black hole binaries}

\begin{figure}
\centering
\includegraphics[width=5.5in,angle=0.]{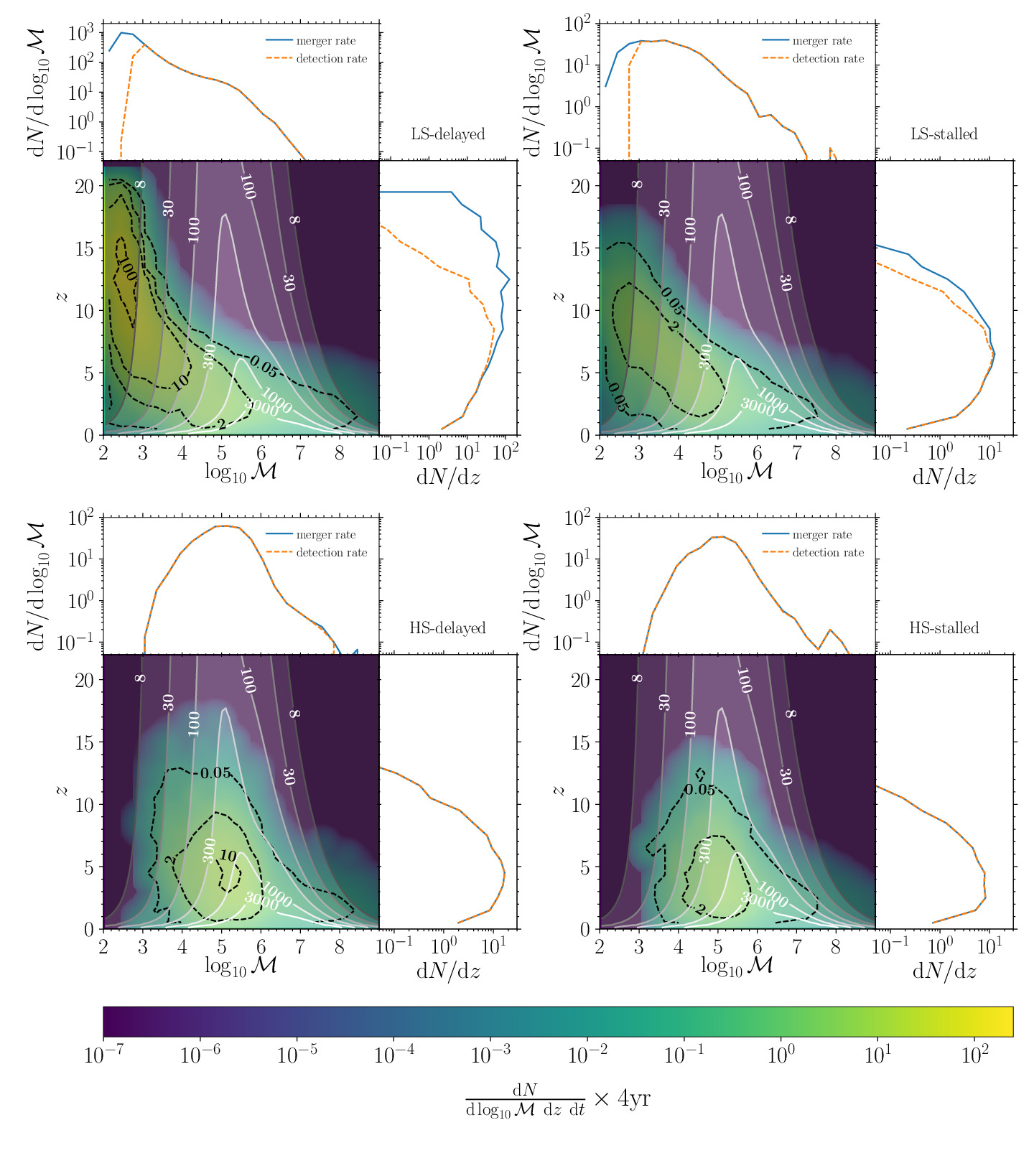}
\caption{LISA observational capabilities vs. predicted MBHB merger
  rates in the chirp mass-redshift plane. In each panel, grey shaded
  contours show the S/N of LISA observations for equal-mass,
  non-spinning binaries. The superimposed yellow-green colour gradient
  with black dashed contours represents the differential number of
  mergers during the planned 4-year mission lifetime. From the
  upper-left panel, clockwise, we show four different astrophysical
  models: LS-delayed, LS-stalled, HS-stalled, and HS-delayed
  \citep[see][for details]{2019MNRAS.486.4044B}. For each model, the
  upper and right-side panels show the merger rate (blue line) and
  detection rate (orange line) distributions marginalized over
  redshift and chirp mass,
  respectively. \citep[From][]{2019MNRAS.486.4044B}}
\label{fig:lisa_smbhb}
\end{figure}

MBHBs are expected to form in large number along the cosmic history \citep{2003ApJ...582..559V}. Pairing in the aftermath of galaxy mergers, they are tracers of structure formation in the Universe, and can be seen by LISA out to $z>20$, beyond the foreseeable capabilities of any electromagnetic (EM) observation. The poor knowledge of protogalaxy and black hole seed formation at high redshift is mirrored in the large uncertainties in detection rate predictions \citep[e.g.][]{2011PhRvD..83d4036S,2020arXiv200603065B}. Nonetheless, LISA is expected to observe between a few and a hundred MBHB coalescences per year. The unique potential of this observatory is shown in Figure~\ref{fig:lisa_smbhb}, where LISA signal-to-noise ratio (S/N) contours for equal-mass, non-spinning binaries are superimposed to the differential distribution of mergers occurring in 4 years (the nominal mission lifetime) in the chirp mass-redshift plane, as predicted by four selected MBH evolution models \citep{2019MNRAS.486.4044B}.

In the case of high-mass seeds from direct collapse shown in the bottom panels of the figure \citep[see][for a recent review]{2018arXiv181012310W}, LISA can see essentially every single merger occurring within the observable Universe. If instead seeds are produced as remnants of popIII stars~\citep{2001ApJ...551L..27M}, as in the top panels of the figure, LISA will miss the first round of mergers, but it will still probe the subsequent growth history of MBHs out to $z\approx 20$. In the latter case, an intriguing possibility is to complement LISA with ground-based 3G observations to fully reconstruct the cosmic history of those systems \citep[see e.g.][]{2020NatAs...4..260J}. In any case, LISA will provide a unique sample of up to several hundred MBHB coalescences: a potential revolution in physics and astronomy.

\subsection{Extracting information}
MBHBs will generally enter the LISA band during their inspiral, completing thousands of cycles before merging within the detector's band. This will allow the accumulation of such high S/N that the main source of error in the parameter recovery, at least for the loudest sources, might come from inaccuracies in the available waveforms rather than from the intrinsic detector noise. In fact, currently available inspiral-merger-ringdown (IMR) waveforms \citep{2017PhRvD..95d4028B,2018arXiv180910113K} are not even close to the needed level of accuracy. This is particular critical for tests of GR with, e.g., ringdown spectroscopy ~\citep{2006PhRvD..73f4030B,2013LRR....16....7G} which relies on measuring tiny deviations from the higher multipoles of the ringdown radiation compared to GR expectations ~\citep{2018PhRvD..97d4048B}, especially to extract information from the higher multipoles of the radiation ~\citep{2019PhRvD..99b4005B}.


Nonetheless, waveforms employed so far include most of the relevant physics and can therefore provide a reliable estimate of LISA's capabilities. As an example, \citep{2016PhRvD..93b4003K} carried a comprehensive study based on spinning precessing post-Newtonian waveforms, corrected for the enhancement in S/N provided by adding merger and ringdown. They found that LISA can recover individual {\it redshifted} masses, i.e. $(1+z)M$, to better than $1\%$ for loud sources at $z<5$. To get the intrinsic mass, however, one must know the redshift to the source, which is computed from the $D_L$ measurement, by assuming a fiducial cosmology. At $z>1$, LISA will measure $D_L$ to a few\% accuracy, and weak lensing will affect the $D_L-z$ conversion adding another few\% error \citep{2010MNRAS.404..858S}. Considering both effects, LISA will provide an estimate of individual source frame masses within $<10$\% relative accuracy for sources at $z<5$. Note that such precise measurements are today available only for MBHs in the local Universe, including SgrA$^*$~\cite{2008ApJ...689.1044G,2009ApJ...692.1075G}, M87~\cite{2019ApJ...875L...1E,2019ApJ...875L...6E}, and few systems powering mega-maser~\cite{1995Natur.373..127M}. The other relevant property of astrophysical MBHs is their spin magnitude and orientation, which are notoriously difficult to measure and are as of today estimated (with large uncertainties) only for $\approx 20$ systems in the low-redshift Universe \citep[see e.g.][]{2014SSRv..183..277R}. Moving to the early epoch of structure formation, estimating parameters of systems at $z>10$ will be more challenging. In particular, the error on $D_L$ tends to become much larger, nonetheless LISA can still place a 95\% lower limit to the source redshift of $\approx 0.66 z$~\citep{2013ASPC..467..103S}.

\subsection{MBH cosmic history reconstruction.}
Because of its excellent parameter estimation capabilities, LISA will deliver an unprecedented catalogue of MBHB coalescences, that will provide precious information about their formation and evolution along the cosmic history \citep{2011PhRvD..83d4036S}. This is because the mass, redshift and spin distribution of LISA events carries the imprint of the underlying physics driving their formation and evolution, including the origin, abundance, mass function and redshift distribution of the first seeds; the detailed properties of the subsequent accretion processes driving their mass growth; the dynamical details of the pairing and hardening process of MBHBs forming in the aftermath of galaxy mergers, and so on. For example, the seeding mechanism as a direct impact on the number of observable sources. Astrophysical low (popIII) and high (direct collapse) mass seed scenarios have been extensively explored and result in very different number of mergers in the LISA band. Furthermore, the MBH seeding process can be connected to the production of primordial BHs in the early Universe \citep{2010RAA....10..495K,2015PhRvD..92b3524C}, a scenario that can be tested by LISA as more quantitative predictions of merger rates become available. On the other hand, measured MBH spins are mainly determined by the geometry of the accretion flow, with prolonged accretion in a defined plane resulting in efficient MBH spin-ups ~\citep{1974ApJ...191..507T}, in contrast to the spin-down caused by interaction with cold gas clouds incoming from random directions ~\citep{2005MNRAS.363...49K}. Mergers also play a role in determining the magnitude and relative orientation of the MBH spins: in gas rich environment, interaction wilt a putative massive circumbinary disk \citep{2009MNRAS.399.2249P} tend to align individual spins with the binary angular momentum, whereas spins of MBHBs merging in gas poor environment are expected to be randomly oriented ~\citep{2007ApJ...661L.147B}. Moreover, the redshift distribution of detected systems is strongly affected by the time required for the binary to complete its journey from kpc scales down to final coalescence, following the host galaxy merger~\citep{2019MNRAS.486.4044B,2020arXiv200603065B}. One of the main challenges of future astrophysical modelling will be to make the best out of the LISA dataset to address the ``inverse problem'' of reconstructing the MBHB cosmic history from observations. In a proof-of-concept study, \cite{2011PhRvD..83d4036S} showed that  LISA can separate different seed models (popIII vs direct collapse) and accretion geometries (coherent vs chaotic), with only an handful of events.

\subsection{EM counterparts and multimessenger astronomy.}
Occurring at the very centre of galaxy merger remnants, MBHBs form and evolve within a dense environment that might favour the presence of EM signals matching the inspiral and coalescence of the pair. As mentioned above, in gas rich environments, binaries are expected to be surrounded by a massive circumbinary disk. Gas can leak from the inner edge of the disc, feeding minidiscs around individual MBHs~\citep{2014ApJ...783..134F}, resulting in a number of distinctive EM signal. For example, feeding of the minidiscs might be modulated over the period of the binary, eventually resulting in a resulting periodicity of their emission ~\citep{2018MNRAS.476.2249T}; the cavity evacuated by the binary torques, removing a significant portion of the inner disc, will produce a distinctive shape of the UV continuum~\citep{2012MNRAS.420..705T}; streams can produce periodic non-thermal X ray bursts upon impact onto the outer edge of the minidiscs ~\citep{2014ApJ...785..115R}; finally, the inverse Compton up-scatter of photons in the corona might produce distinctive double K$\alpha$ lines~\citep{2012MNRAS.420..860S}.The main challenge will be the detection and identification of all those putative features. Being an omni-directional detector, LISA sky localization capabilities are mostly determined by the evolution of the antenna response function as it move along its orbit around the Sun. For MBHBs this will allow localization of $z<2$ sources within $\Delta{\Omega}<10(0.5)\,$deg$^2$ weeks(hours) before coalescence~\citep{2010PhRvD..81f4014M,2020arXiv200612513M}. This is a remarkable feat foe GW astronomy, allowing for  searches with optical, radio and X-ray wide-field instruments, such as LSST \citep{2009arXiv0912.0201L}, SKA and Athena \citep{2020NatAs...4...26M}. After merger, the high S/N added at coalescence, LISA sky localization will improve to several arcmin$^2$; deeper EM  observations might then reveal a number of features related to the post-merger dynamics of the surrounding medium. These include the the birth of a quasar as the gas in the circumbinary disc refill the cavity and is efficiently accreted~\citep{2005ApJ...622L..93M}, the launch of a relativistic jet~\citep{2010Sci...329..927P}, or non thermal emission from shocks prompted within the disk by the sudden change of the potential due to gravitational recoil~\citep{2010MNRAS.401.2021R}. Convincing identification of any such counterpart would be an unprecedented milestone in accretion physics, opening up the study of interaction between gravity and matter in the time-dependent, strong field regime of a merging binary, as well as probing accretion onto MBHs of known masses and spins thus allowing, among other things, to test theoretical conjectures linking MBH spins to jet launching~\citep{1977MNRAS.179..433B}. Last but not least, joint EM and GW detections of MBHB will provide a unique class of standard sirens, extending up to $z>5$~\citep{2016JCAP...04..002T}, thus probing the expansion history of the Universe in uncharted territory.

\section{Extreme mass ratio inspirals}

EMRIs are distinct from MBHBs both in their properties and their origin. As the name indicates, they are binaries involving objects of very different masses, generally a MBH interacting with a CO that can be a WD, NS or stellar mass BH. Consequently, their origin is not related to galaxy mergers or, more broadly, to the hierarchical structure formation paradigm, but is rooted in the relativistic dynamics of dense nuclei. Sitting at galactic centres,in fact, MBHs are surrounded by a dense distribution of stars and COs. In such a dense environment, the central MBH can 'capture' a stellar BH as a result of several dynamical processes, including different flavour of relaxation mechanisms deflecting BHs onto low angular momentum orbits or the tidal breakup of a compact binary close to the MBH. The captured BH will then inspiral onto the central MBH completing millions of orbits before eventually plunging into it~\citep{2018LRR....21....4A}. The detection of the resulting GW signal poses a major challenge for GW modellers, since it requires matching hundred of thousands of cycles with accurate enough waveform templates \citep{2004PhRvD..69h2005B,2009CQGra..26u3001B,2015CQGra..32w2002C,2017PhRvD..96d4005C}. But payouts are well worth the investment of theoretical and computational resources. Upon detection EMRIs will deliver unprecedented measurements of the system parameters, including the central MBH mass and spin to a precision of $<10^{-4}$, a luminosity distance accuracy of a few percent, and sky localization within $\approx 1\,$deg$^2$~\citep{2004PhRvD..69h2005B,2017PhRvD..95j3012B}, making them formidable probes of MBH astrophysics, fundamental physics and cosmology.

Capable of detecting EMRIs out to $z \approx 2$, LISA will detect from few to thousands of these systems per year ~\citep{2017PhRvD..95j3012B}. Very uncertain rates stem from poorly known underlying physics, meaning that EMRIs will provide a new wealth of information about the conditions of dense nuclei, in particular the mass function and occupation fractions of dormant MBHs in the mass range $10^5\msun-10^6\msun$, difficult to probe by other means~\citep{2010PhRvD..81j4014G}. Source abundance and individual EMRI parameters, such as eccentricity and orbital inclination, will help constrain their formation channel, shedding new light on extreme dynamics in dense nuclei~\citep{2018LRR....21....4A}. A fraction of EMRIs might also form and evolve within AGN discs~\citep{2007MNRAS.374..515L}. If this is the case, drag from the disc will leave distinctive signatures in the waveform, giving us access to the conditions of the plasma in the mid-plane of optically thick accretion discs, something that is beyond the reach of photon-based astronomy ~\citep{2011PhRvD..84b4032K,2014PhRvD..89j4059B}. Exquisite parameter estimation accuracy makes EMRIs unique tools for probing space-time. For example the central MBH quadrupole moment can be measured to a fractional precision of $<10^{-4}$, allowing the detection of tiny deviation from Kerr geometry. Finally, although generally lacking EM counterparts, the excellent measurement of EMRIs distance and sky location  will allow to effectively determine their redshift via statistical methods. Estimates suggest that $H_0$ could be measured to an accuracy of $\approx 1$\% with an ensemble of 20 EMRIs detected out to $z\approx 0.5$~\citep{2008PhRvD..77d3512M}.

\section{Stellar-mass black hole binaries and multiband detections}
Last but not least, LISA will observe stellar-mass BHBs still far from coalescence, before they enter the ground-based detector band. This was soon realized after the detection of GW150914, a system so massive and nearby that would have been observed by LISA with S/N $\approx 5$ about five years before coalescence~\citep{2016PhRvL.116w1102S}. Subsequent studies have demonstrated that LISA can detect several tens of BHBs, up to hundreds of years before coalescence. A fraction of them will be caught in the last few years of inspiral, and will cross all the way to the LIGO-Virgo band, paving the way to multiband GW astronomy~\citep{2016MNRAS.462.2177K,2017JPhCS.840a2018S,2019PhRvD..99j3004G}. LISA will localize these multiband sources within $\approx 0.1\,$deg$^2$, predicting their coalescence time with an error of  $<10\,$s. We will therefore be in the unprecedented position of knowing exactly where and when a BHB is going to merge, a condition that will allow pre-pointing of EM facilities to search for a possible counterparts {\it coincident} with the merger with a depth which is inconceivable with wide-field monitors~\citep{2016PhRvL.116w1102S}. 
Reconstructing the phase evolution of the system across five decades in frequency, and possibly fine-tuning the sensitivity of Earth-based detectors, will lead to improved tests of general relativity~\cite{2016PhRvL.116x1104B,2019arXiv190513155C,2019BAAS...51c..32B,2017PhRvD..96h4039C,2018arXiv180700075T,2019arXiv190513460G}. As an example, observations of the same source in he early and late inspiral will place unique constraints on additional emission multipoles \cite{2016PhRvL.116x1104B}.

Even without multiband observations, detecting stellar-origin BHBs with LISA may have important astrophysical implications. Far from coalescence, LISA can measure the eccentricity $e$ of these binaries as long as $e\gtrsim 10^{-3}$ at GW frequencies $f\sim 10^{-2}$~Hz~\citep{2016PhRvD..94f4020N}. Field binaries are expected to have small eccentricities at these frequencies~\citep{2011A&A...527A..70K}, therefore these measurements can be used to discriminate between the dynamical and field formation channels~\citep{2016ApJ...830L..18B,2017MNRAS.465.4375N,2018MNRAS.481.5445S}.
Combined with ground-based spin measurements, LISA eccentricity measurements can have an important role in our understanding of BHB formation. If the rate turns out to be large, specific stellar sub-populations could potentially be constrained \citep[e.g.][]{2019PhRvD..99j3004G}.
Cosmology will also benefit. Similar to EMRIs, the sky location and distance of at least a subset of these systems can be precise enough that we could use them as standard candles, allowing for an independent statistical measurement of $H_0$ within a few percent accuracy~\citep{2017PhRvD..95h3525K,2018MNRAS.475.3485D}.


\section{Conclusions}
The future of GW astronomy is going to be loud. Building on the successes of LIGO and Virgo, the GW community is investing in a number of projects that will tremendously expand our knowledge of the dark side of the Universe. 3G ground based detectors will observe hundreds of thousands CO mergers across the Universe and PTAs will unveil the most massive black hole binaries in the Universe. In this context, LISA will be one of our finest ears on the Universe. By surveying the milli-Hz frequency band, LISA will detect a variety of GW sources, across several decades in the mass scale, from the Solar neighbourhood back to the formation of the first cosmic structure, promising an unprecedented revolution in our understanding of the Universe.






\section*{Acknowledgements}
The Author is supported by the European Research Council (ERC) under the European Union’s Horizon 2020 research and innovation program ERC-2018-COG under grant
agreement No 818691 (B Massive). The Author is also indebited to Emanuele Berti for early contributions to this review.

\bibliographystyle{frontiersinSCNS_ENG_HUMS} 
\bibliography{biblio}

\begin{thebibliography}{96}
\providecommand{\natexlab}[1]{#1}
\expandafter\ifx\csname urlstyle\endcsname\relax
  \providecommand{\doi}[1]{doi:\discretionary{}{}{}#1}\else
  \providecommand{\doi}{doi:\discretionary{}{}{}\begingroup
  \urlstyle{rm}\Url}\fi
\providecommand{\selectlanguage}[1]{\relax}
\providecommand{\bibAnnoteFile}[1]{%
  \IfFileExists{#1}{\begin{quotation}\noindent\textsc{Key:} #1\\
  \textsc{Annotation:}\ \input{#1}\end{quotation}}{}}
\providecommand{\bibAnnote}[2]{%
  \begin{quotation}\noindent\textsc{Key:} #1\\
  \textsc{Annotation:}\ #2\end{quotation}}

\bibitem[{{Abbott} et~al.(2009){Abbott}, {Abbott}, {Adhikari}, {Ajith},
  {Allen}, {Allen} et~al.}]{2009RPPh...72g6901A}
{Abbott}, B.~P., {Abbott}, R., {Adhikari}, R., {Ajith}, P., {Allen}, B.,
  {Allen}, G., et~al. (2009).
\newblock {LIGO: the Laser Interferometer Gravitational-Wave Observatory}.
\newblock \emph{Reports on Progress in Physics} 72, 076901.
\newblock \doi{10.1088/0034-4885/72/7/076901}
\bibAnnoteFile{2009RPPh...72g6901A}

\bibitem[{{Abbott} et~al.(2019){Abbott}, {Abbott}, and
  {Adhikari}}]{2019PhRvX...9c1040A}
{Abbott}, B.~P., {Abbott}, R., and {Adhikari}, R. e.~a. (2019).
\newblock {GWTC-1: A Gravitational-Wave Transient Catalog of Compact Binary
  Mergers Observed by LIGO and Virgo during the First and Second Observing
  Runs}.
\newblock \emph{Physical Review X} 9, 031040.
\newblock \doi{10.1103/PhysRevX.9.031040}
\bibAnnoteFile{2019PhRvX...9c1040A}

\bibitem[{{Acernese} et~al.(2015){Acernese}, {Agathos}, {Agatsuma}, {Aisa},
  {Allemandou}, {Allocca} et~al.}]{2015CQGra..32b4001A}
{Acernese}, F., {Agathos}, M., {Agatsuma}, K., {Aisa}, D., {Allemandou}, N.,
  {Allocca}, A., et~al. (2015).
\newblock {Advanced Virgo: a second-generation interferometric gravitational
  wave detector}.
\newblock \emph{Classical and Quantum Gravity} 32, 024001.
\newblock \doi{10.1088/0264-9381/32/2/024001}
\bibAnnoteFile{2015CQGra..32b4001A}

\bibitem[{{Amaro-Seoane}(2018)}]{2018LRR....21....4A}
{Amaro-Seoane}, P. (2018).
\newblock {Relativistic dynamics and extreme mass ratio inspirals}.
\newblock \emph{Living Reviews in Relativity} 21, 4.
\newblock \doi{10.1007/s41114-018-0013-8}
\bibAnnoteFile{2018LRR....21....4A}

\bibitem[{{Amaro-Seoane}(2019)}]{2019PhRvD..99l3025A}
{Amaro-Seoane}, P. (2019).
\newblock {Extremely large mass-ratio inspirals}.
\newblock \emph{Phys. Rev. D} 99, 123025.
\newblock \doi{10.1103/PhysRevD.99.123025}
\bibAnnoteFile{2019PhRvD..99l3025A}

\bibitem[{{Amaro-Seoane} et~al.(2017){Amaro-Seoane}, {Audley}, {Babak},
  {Baker}, {Barausse}, {Bender} et~al.}]{2017arXiv170200786A}
{Amaro-Seoane}, P., {Audley}, H., {Babak}, S., {Baker}, J., {Barausse}, E.,
  {Bender}, P., et~al. (2017).
\newblock {Laser Interferometer Space Antenna}.
\newblock \emph{arXiv e-prints}
\bibAnnoteFile{2017arXiv170200786A}

\bibitem[{{Arzoumanian} et~al.(2018){Arzoumanian}, {Brazier}, {Burke-Spolaor},
  {Chamberlin}, {Chatterjee}, {Christy} et~al.}]{2018ApJS..235...37A}
{Arzoumanian}, Z., {Brazier}, A., {Burke-Spolaor}, S., {Chamberlin}, S.,
  {Chatterjee}, S., {Christy}, B., et~al. (2018).
\newblock {The NANOGrav 11-year Data Set: High-precision Timing of 45
  Millisecond Pulsars}.
\newblock \emph{Astrophys. J. Supplements} 235, 37.
\newblock \doi{10.3847/1538-4365/aab5b0}
\bibAnnoteFile{2018ApJS..235...37A}

\bibitem[{{Babak} et~al.(2017){Babak}, {Gair}, {Sesana}, {Barausse},
  {Sopuerta}, {Berry} et~al.}]{2017PhRvD..95j3012B}
{Babak}, S., {Gair}, J., {Sesana}, A., {Barausse}, E., {Sopuerta}, C.~F.,
  {Berry}, C.~P.~L., et~al. (2017).
\newblock {Science with the space-based interferometer LISA. V. Extreme
  mass-ratio inspirals}.
\newblock \emph{Phys. Rev. D} 95, 103012.
\newblock \doi{10.1103/PhysRevD.95.103012}
\bibAnnoteFile{2017PhRvD..95j3012B}

\bibitem[{{Baibhav} and {Berti}(2019)}]{2019PhRvD..99b4005B}
{Baibhav}, V. and {Berti}, E. (2019).
\newblock {Multimode black hole spectroscopy}.
\newblock \emph{Phys. Rev. D} 99, 024005.
\newblock \doi{10.1103/PhysRevD.99.024005}
\bibAnnoteFile{2019PhRvD..99b4005B}

\bibitem[{{Baibhav} et~al.(2018){Baibhav}, {Berti}, {Cardoso}, and
  {Khanna}}]{2018PhRvD..97d4048B}
{Baibhav}, V., {Berti}, E., {Cardoso}, V., and {Khanna}, G. (2018).
\newblock {Black hole spectroscopy: Systematic errors and ringdown energy
  estimates}.
\newblock \emph{Phys. Rev. D} 97, 044048.
\newblock \doi{10.1103/PhysRevD.97.044048}
\bibAnnoteFile{2018PhRvD..97d4048B}

\bibitem[{{Barack}(2009)}]{2009CQGra..26u3001B}
{Barack}, L. (2009).
\newblock {TOPICAL REVIEW: Gravitational self-force in extreme mass-ratio
  inspirals}.
\newblock \emph{Classical and Quantum Gravity} 26, 213001.
\newblock \doi{10.1088/0264-9381/26/21/213001}
\bibAnnoteFile{2009CQGra..26u3001B}

\bibitem[{{Barack} and {Cutler}(2004)}]{2004PhRvD..69h2005B}
{Barack}, L. and {Cutler}, C. (2004).
\newblock {LISA capture sources: Approximate waveforms, signal-to-noise ratios,
  and parameter estimation accuracy}.
\newblock \emph{Phys. Rev. D} 69, 082005.
\newblock \doi{10.1103/PhysRevD.69.082005}
\bibAnnoteFile{2004PhRvD..69h2005B}

\bibitem[{{Barausse} et~al.(2014){Barausse}, {Cardoso}, and
  {Pani}}]{2014PhRvD..89j4059B}
{Barausse}, E., {Cardoso}, V., and {Pani}, P. (2014).
\newblock {Can environmental effects spoil precision gravitational-wave
  astrophysics?}
\newblock \emph{Phys. Rev. D} 89, 104059.
\newblock \doi{10.1103/PhysRevD.89.104059}
\bibAnnoteFile{2014PhRvD..89j4059B}

\bibitem[{{Barausse} et~al.(2020){Barausse}, {Dvorkin}, {Tremmel}, {Volonteri},
  and {Bonetti}}]{2020arXiv200603065B}
{Barausse}, E., {Dvorkin}, I., {Tremmel}, M., {Volonteri}, M., and {Bonetti},
  M. (2020).
\newblock {Massive black hole merger rates: the effect of kpc separation
  wandering and supernova feedback}.
\newblock \emph{arXiv e-prints} , arXiv:2006.03065
\bibAnnoteFile{2020arXiv200603065B}

\bibitem[{{Barausse} et~al.(2016){Barausse}, {Yunes}, and
  {Chamberlain}}]{2016PhRvL.116x1104B}
{Barausse}, E., {Yunes}, N., and {Chamberlain}, K. (2016).
\newblock {Theory-Agnostic Constraints on Black-Hole Dipole Radiation with
  Multiband Gravitational-Wave Astrophysics}.
\newblock \emph{Physical Review Letters} 116, 241104.
\newblock \doi{10.1103/PhysRevLett.116.241104}
\bibAnnoteFile{2016PhRvL.116x1104B}

\bibitem[{{Berti} et~al.(2019){Berti}, {Barausse}, {Cholis}, {Garcia-Bellido},
  {Holley-Bockelmann}, {Hughes} et~al.}]{2019BAAS...51c..32B}
{Berti}, E., {Barausse}, E., {Cholis}, I., {Garcia-Bellido}, J.,
  {Holley-Bockelmann}, K., {Hughes}, S.~A., et~al. (2019).
\newblock {Tests of General Relativity and Fundamental Physics with Space-based
  Gravitational Wave Detectors}.
\newblock In \emph{Bulletin of the American Astronomical Society}. vol.~51 of
  \emph{Bulletin of the American Astron. Soc.}, 32
\bibAnnoteFile{2019BAAS...51c..32B}

\bibitem[{{Berti} et~al.(2006){Berti}, {Cardoso}, and
  {Will}}]{2006PhRvD..73f4030B}
{Berti}, E., {Cardoso}, V., and {Will}, C.~M. (2006).
\newblock {Gravitational-wave spectroscopy of massive black holes with the
  space interferometer LISA}.
\newblock \emph{Phys. Rev. D} 73, 064030.
\newblock \doi{10.1103/PhysRevD.73.064030}
\bibAnnoteFile{2006PhRvD..73f4030B}

\bibitem[{{Blandford} and {Znajek}(1977)}]{1977MNRAS.179..433B}
{Blandford}, R.~D. and {Znajek}, R.~L. (1977).
\newblock {Electromagnetic extraction of energy from Kerr black holes}.
\newblock \emph{Mon. Not. R. Ast. Soc.} 179, 433--456.
\newblock \doi{10.1093/mnras/179.3.433}
\bibAnnoteFile{1977MNRAS.179..433B}

\bibitem[{{Bogdanovi{\'c}} et~al.(2007){Bogdanovi{\'c}}, {Reynolds}, and
  {Miller}}]{2007ApJ...661L.147B}
{Bogdanovi{\'c}}, T., {Reynolds}, C.~S., and {Miller}, M.~C. (2007).
\newblock {Alignment of the Spins of Supermassive Black Holes Prior to
  Coalescence}.
\newblock \emph{Astrophys. J. Lett.} 661, L147--L150.
\newblock \doi{10.1086/518769}
\bibAnnoteFile{2007ApJ...661L.147B}

\bibitem[{{Boh{\'e}} et~al.(2017){Boh{\'e}}, {Shao}, {Taracchini}, {Buonanno},
  {Babak}, {Harry} et~al.}]{2017PhRvD..95d4028B}
{Boh{\'e}}, A., {Shao}, L., {Taracchini}, A., {Buonanno}, A., {Babak}, S.,
  {Harry}, I.~W., et~al. (2017).
\newblock {Improved effective-one-body model of spinning, nonprecessing binary
  black holes for the era of gravitational-wave astrophysics with advanced
  detectors}.
\newblock \emph{Phys. Rev. D} 95, 044028.
\newblock \doi{10.1103/PhysRevD.95.044028}
\bibAnnoteFile{2017PhRvD..95d4028B}

\bibitem[{{Bonetti} et~al.(2019){Bonetti}, {Sesana}, {Haardt}, {Barausse}, and
  {Colpi}}]{2019MNRAS.486.4044B}
{Bonetti}, M., {Sesana}, A., {Haardt}, F., {Barausse}, E., and {Colpi}, M.
  (2019).
\newblock {Post-Newtonian evolution of massive black hole triplets in galactic
  nuclei - IV. Implications for LISA}.
\newblock \emph{Mon. Not. R. Ast. Soc.} 486, 4044--4060.
\newblock \doi{10.1093/mnras/stz903}
\bibAnnoteFile{2019MNRAS.486.4044B}

\bibitem[{{Breivik} et~al.(2016){Breivik}, {Rodriguez}, {Larson}, {Kalogera},
  and {Rasio}}]{2016ApJ...830L..18B}
{Breivik}, K., {Rodriguez}, C.~L., {Larson}, S.~L., {Kalogera}, V., and
  {Rasio}, F.~A. (2016).
\newblock {Distinguishing between Formation Channels for Binary Black Holes
  with LISA}.
\newblock \emph{Astrophys. J. Lett.} 830, L18.
\newblock \doi{10.3847/2041-8205/830/1/L18}
\bibAnnoteFile{2016ApJ...830L..18B}

\bibitem[{{Caprini} and {Figueroa}(2018)}]{2018CQGra..35p3001C}
{Caprini}, C. and {Figueroa}, D.~G. (2018).
\newblock {Cosmological backgrounds of gravitational waves}.
\newblock \emph{Classical and Quantum Gravity} 35, 163001.
\newblock \doi{10.1088/1361-6382/aac608}
\bibAnnoteFile{2018CQGra..35p3001C}

\bibitem[{{Carson} and {Yagi}(2019)}]{2019arXiv190513155C}
{Carson}, Z. and {Yagi}, K. (2019).
\newblock {Multi-band gravitational wave tests of general relativity}.
\newblock \emph{arXiv e-prints}
\bibAnnoteFile{2019arXiv190513155C}

\bibitem[{{Chamberlain} and {Yunes}(2017)}]{2017PhRvD..96h4039C}
{Chamberlain}, K. and {Yunes}, N. (2017).
\newblock {Theoretical physics implications of gravitational wave observation
  with future detectors}.
\newblock \emph{Phys. Rev. D} 96, 084039.
\newblock \doi{10.1103/PhysRevD.96.084039}
\bibAnnoteFile{2017PhRvD..96h4039C}

\bibitem[{{Chua} and {Gair}(2015)}]{2015CQGra..32w2002C}
{Chua}, A.~J.~K. and {Gair}, J.~R. (2015).
\newblock {Improved analytic extreme-mass-ratio inspiral model for scoping out
  eLISA data analysis}.
\newblock \emph{Classical and Quantum Gravity} 32, 232002.
\newblock \doi{10.1088/0264-9381/32/23/232002}
\bibAnnoteFile{2015CQGra..32w2002C}

\bibitem[{{Chua} et~al.(2017){Chua}, {Moore}, and {Gair}}]{2017PhRvD..96d4005C}
{Chua}, A.~J.~K., {Moore}, C.~J., and {Gair}, J.~R. (2017).
\newblock {Augmented kludge waveforms for detecting extreme-mass-ratio
  inspirals}.
\newblock \emph{Phys. Rev. D} 96, 044005.
\newblock \doi{10.1103/PhysRevD.96.044005}
\bibAnnoteFile{2017PhRvD..96d4005C}

\bibitem[{{Clesse} and {Garc{\'\i}a-Bellido}(2015)}]{2015PhRvD..92b3524C}
{Clesse}, S. and {Garc{\'\i}a-Bellido}, J. (2015).
\newblock {Massive primordial black holes from hybrid inflation as dark matter
  and the seeds of galaxies}.
\newblock \emph{Phys. Rev. D.} 92, 023524.
\newblock \doi{10.1103/PhysRevD.92.023524}
\bibAnnoteFile{2015PhRvD..92b3524C}

\bibitem[{{Del Pozzo} et~al.(2018){Del Pozzo}, {Sesana}, and
  {Klein}}]{2018MNRAS.475.3485D}
{Del Pozzo}, W., {Sesana}, A., and {Klein}, A. (2018).
\newblock {Stellar binary black holes in the LISA band: a new class of standard
  sirens}.
\newblock \emph{Mon. Not. R. Ast. Soc.} 475, 3485--3492.
\newblock \doi{10.1093/mnras/sty057}
\bibAnnoteFile{2018MNRAS.475.3485D}

\bibitem[{{Desvignes} et~al.(2016){Desvignes}, {Caballero}, {Lentati},
  {Verbiest}, {Champion}, {Stappers} et~al.}]{2016MNRAS.458.3341D}
{Desvignes}, G., {Caballero}, R.~N., {Lentati}, L., {Verbiest}, J.~P.~W.,
  {Champion}, D.~J., {Stappers}, B.~W., et~al. (2016).
\newblock {High-precision timing of 42 millisecond pulsars with the European
  Pulsar Timing Array}.
\newblock \emph{Mon. Not. R. Ast. Soc.} 458, 3341--3380.
\newblock \doi{10.1093/mnras/stw483}
\bibAnnoteFile{2016MNRAS.458.3341D}

\bibitem[{{Dewdney} et~al.(2009){Dewdney}, {Hall}, {Schilizzi}, and
  {Lazio}}]{2009IEEEP..97.1482D}
{Dewdney}, P.~E., {Hall}, P.~J., {Schilizzi}, R.~T., and {Lazio}, T.~J.~L.~W.
  (2009).
\newblock {The Square Kilometre Array}.
\newblock \emph{IEEE Proceedings} 97, 1482--1496.
\newblock \doi{10.1109/JPROC.2009.2021005}
\bibAnnoteFile{2009IEEEP..97.1482D}

\bibitem[{{Event Horizon Telescope Collaboration}
  et~al.(2019{\natexlab{a}}){Event Horizon Telescope Collaboration}, {Akiyama},
  {Alberdi}, {Alef}, {Asada}, {Azulay} et~al.}]{2019ApJ...875L...1E}
{Event Horizon Telescope Collaboration}, {Akiyama}, K., {Alberdi}, A., {Alef},
  W., {Asada}, K., {Azulay}, R., et~al. (2019{\natexlab{a}}).
\newblock {First M87 Event Horizon Telescope Results. I. The Shadow of the
  Supermassive Black Hole}.
\newblock \emph{Astrophys. J. Lett.} 875, L1.
\newblock \doi{10.3847/2041-8213/ab0ec7}
\bibAnnoteFile{2019ApJ...875L...1E}

\bibitem[{{Event Horizon Telescope Collaboration}
  et~al.(2019{\natexlab{b}}){Event Horizon Telescope Collaboration}, {Akiyama},
  {Alberdi}, {Alef}, {Asada}, {Azulay} et~al.}]{2019ApJ...875L...6E}
{Event Horizon Telescope Collaboration}, {Akiyama}, K., {Alberdi}, A., {Alef},
  W., {Asada}, K., {Azulay}, R., et~al. (2019{\natexlab{b}}).
\newblock {First M87 Event Horizon Telescope Results. VI. The Shadow and Mass
  of the Central Black Hole}.
\newblock \emph{Astrophys. J. Lett.} 875, L6.
\newblock \doi{10.3847/2041-8213/ab1141}
\bibAnnoteFile{2019ApJ...875L...6E}

\bibitem[{{Farris} et~al.(2014){Farris}, {Duffell}, {MacFadyen}, and
  {Haiman}}]{2014ApJ...783..134F}
{Farris}, B.~D., {Duffell}, P., {MacFadyen}, A.~I., and {Haiman}, Z. (2014).
\newblock {Binary Black Hole Accretion from a Circumbinary Disk: Gas Dynamics
  inside the Central Cavity}.
\newblock \emph{Astrophys. J.} 783, 134.
\newblock \doi{10.1088/0004-637X/783/2/134}
\bibAnnoteFile{2014ApJ...783..134F}

\bibitem[{{Foster} and {Backer}(1990)}]{1990ApJ...361..300F}
{Foster}, R.~S. and {Backer}, D.~C. (1990).
\newblock {Constructing a Pulsar Timing Array}.
\newblock \emph{Astrophys. J.} 361, 300.
\newblock \doi{10.1086/169195}
\bibAnnoteFile{1990ApJ...361..300F}

\bibitem[{{Freitag}(2003)}]{2003ApJ...583L..21F}
{Freitag}, M. (2003).
\newblock {Gravitational Waves from Stars Orbiting the Sagittarius A* Black
  Hole}.
\newblock \emph{Astrphys. J. Lett.} 583, L21--L24.
\newblock \doi{10.1086/367813}
\bibAnnoteFile{2003ApJ...583L..21F}

\bibitem[{{Gair} et~al.(2010){Gair}, {Tang}, and
  {Volonteri}}]{2010PhRvD..81j4014G}
{Gair}, J.~R., {Tang}, C., and {Volonteri}, M. (2010).
\newblock {LISA extreme-mass-ratio inspiral events as probes of the black hole
  mass function}.
\newblock \emph{Phys. Rev. D} 81, 104014.
\newblock \doi{10.1103/PhysRevD.81.104014}
\bibAnnoteFile{2010PhRvD..81j4014G}

\bibitem[{{Gair} et~al.(2013){Gair}, {Vallisneri}, {Larson}, and
  {Baker}}]{2013LRR....16....7G}
{Gair}, J.~R., {Vallisneri}, M., {Larson}, S.~L., and {Baker}, J.~G. (2013).
\newblock {Testing General Relativity with Low-Frequency, Space-Based
  Gravitational-Wave Detectors}.
\newblock \emph{Living Reviews in Relativity} 16, 7.
\newblock \doi{10.12942/lrr-2013-7}
\bibAnnoteFile{2013LRR....16....7G}

\bibitem[{{Gerosa} et~al.(2019){Gerosa}, {Ma}, {Wong}, {Berti},
  {O'Shaughnessy}, {Chen} et~al.}]{2019PhRvD..99j3004G}
{Gerosa}, D., {Ma}, S., {Wong}, K. W.~K., {Berti}, E., {O'Shaughnessy}, R.,
  {Chen}, Y., et~al. (2019).
\newblock {Multiband gravitational-wave event rates and stellar physics}.
\newblock \emph{Phys. Rev. D.} 99, 103004.
\newblock \doi{10.1103/PhysRevD.99.103004}
\bibAnnoteFile{2019PhRvD..99j3004G}

\bibitem[{{Ghez} et~al.(2008){Ghez}, {Salim}, {Weinberg}, {Lu}, {Do}, {Dunn}
  et~al.}]{2008ApJ...689.1044G}
{Ghez}, A.~M., {Salim}, S., {Weinberg}, N.~N., {Lu}, J.~R., {Do}, T., {Dunn},
  J.~K., et~al. (2008).
\newblock {Measuring Distance and Properties of the Milky Way's Central
  Supermassive Black Hole with Stellar Orbits}.
\newblock \emph{Astrophys. J.} 689, 1044--1062.
\newblock \doi{10.1086/592738}
\bibAnnoteFile{2008ApJ...689.1044G}

\bibitem[{{Gillessen} et~al.(2009){Gillessen}, {Eisenhauer}, {Trippe},
  {Alexander}, {Genzel}, {Martins} et~al.}]{2009ApJ...692.1075G}
{Gillessen}, S., {Eisenhauer}, F., {Trippe}, S., {Alexander}, T., {Genzel}, R.,
  {Martins}, F., et~al. (2009).
\newblock {Monitoring Stellar Orbits Around the Massive Black Hole in the
  Galactic Center}.
\newblock \emph{Astrophys. J.} 692, 1075--1109.
\newblock \doi{10.1088/0004-637X/692/2/1075}
\bibAnnoteFile{2009ApJ...692.1075G}

\bibitem[{{Gnocchi} et~al.(2019){Gnocchi}, {Maselli}, {Abdelsalhin},
  {Giacobbo}, and {Mapelli}}]{2019arXiv190513460G}
{Gnocchi}, G., {Maselli}, A., {Abdelsalhin}, T., {Giacobbo}, N., and {Mapelli},
  M. (2019).
\newblock {Bounding Alternative Theories of Gravity with Multi-Band GW
  Observations}.
\newblock \emph{arXiv e-prints}
\bibAnnoteFile{2019arXiv190513460G}

\bibitem[{{Jani} et~al.(2019){Jani}, {Shoemaker}, and
  {Cutler}}]{2020NatAs...4..260J}
{Jani}, K., {Shoemaker}, D., and {Cutler}, C. (2019).
\newblock {Detectability of intermediate-mass black holes in multiband
  gravitational wave astronomy}.
\newblock \emph{Nature Astronomy} 4, 260--265.
\newblock \doi{10.1038/s41550-019-0932-7}
\bibAnnoteFile{2020NatAs...4..260J}

\bibitem[{{Kelley} et~al.(2018){Kelley}, {Blecha}, {Hernquist}, {Sesana}, and
  {Taylor}}]{2018MNRAS.477..964K}
{Kelley}, L.~Z., {Blecha}, L., {Hernquist}, L., {Sesana}, A., and {Taylor},
  S.~R. (2018).
\newblock {Single sources in the low-frequency gravitational wave sky:
  properties and time to detection by pulsar timing arrays}.
\newblock \emph{Mon. Not. R. Ast. Soc.} 477, 964--976.
\newblock \doi{10.1093/mnras/sty689}
\bibAnnoteFile{2018MNRAS.477..964K}

\bibitem[{{Kerr} et~al.(2020){Kerr}, {Reardon}, {Hobbs}, {Shannon},
  {Manchester}, {Dai} et~al.}]{2020PASA...37...20K}
{Kerr}, M., {Reardon}, D.~J., {Hobbs}, G., {Shannon}, R.~M., {Manchester},
  R.~N., {Dai}, S., et~al. (2020).
\newblock {The Parkes Pulsar Timing Array project: second data release}.
\newblock \emph{PASA} 37, e020.
\newblock \doi{10.1017/pasa.2020.11}
\bibAnnoteFile{2020PASA...37...20K}

\bibitem[{{Khan} et~al.(2018){Khan}, {Chatziioannou}, {Hannam}, and
  {Ohme}}]{2018arXiv180910113K}
{Khan}, S., {Chatziioannou}, K., {Hannam}, M., and {Ohme}, F. (2018).
\newblock {Phenomenological model for the gravitational-wave signal from
  precessing binary black holes with two-spin effects}.
\newblock \emph{arXiv e-prints}
\bibAnnoteFile{2018arXiv180910113K}

\bibitem[{{Khlopov}(2010)}]{2010RAA....10..495K}
{Khlopov}, M.~Y. (2010).
\newblock {Primordial black holes}.
\newblock \emph{Research in Astronomy and Astrophysics} 10, 495--528.
\newblock \doi{10.1088/1674-4527/10/6/001}
\bibAnnoteFile{2010RAA....10..495K}

\bibitem[{{King} et~al.(2005){King}, {Lubow}, {Ogilvie}, and
  {Pringle}}]{2005MNRAS.363...49K}
{King}, A.~R., {Lubow}, S.~H., {Ogilvie}, G.~I., and {Pringle}, J.~E. (2005).
\newblock {Aligning spinning black holes and accretion discs}.
\newblock \emph{Mon. Not. R. Ast. Soc.} 363, 49--56.
\newblock \doi{10.1111/j.1365-2966.2005.09378.x}
\bibAnnoteFile{2005MNRAS.363...49K}

\bibitem[{{Klein} et~al.(2016){Klein}, {Barausse}, {Sesana}, {Petiteau},
  {Berti}, {Babak} et~al.}]{2016PhRvD..93b4003K}
{Klein}, A., {Barausse}, E., {Sesana}, A., {Petiteau}, A., {Berti}, E.,
  {Babak}, S., et~al. (2016).
\newblock {Science with the space-based interferometer eLISA: Supermassive
  black hole binaries}.
\newblock \emph{Phys. Rev. D} 93, 024003.
\newblock \doi{10.1103/PhysRevD.93.024003}
\bibAnnoteFile{2016PhRvD..93b4003K}

\bibitem[{{Kocsis} et~al.(2011){Kocsis}, {Yunes}, and
  {Loeb}}]{2011PhRvD..84b4032K}
{Kocsis}, B., {Yunes}, N., and {Loeb}, A. (2011).
\newblock {Observable signatures of extreme mass-ratio inspiral black hole
  binaries embedded in thin accretion disks}.
\newblock \emph{Phys. Rev. D} 84, 024032.
\newblock \doi{10.1103/PhysRevD.84.024032}
\bibAnnoteFile{2011PhRvD..84b4032K}

\bibitem[{{Kowalska} et~al.(2011){Kowalska}, {Bulik}, {Belczynski}, {Dominik},
  and {Gondek-Rosinska}}]{2011A&A...527A..70K}
{Kowalska}, I., {Bulik}, T., {Belczynski}, K., {Dominik}, M., and
  {Gondek-Rosinska}, D. (2011).
\newblock {The eccentricity distribution of compact binaries}.
\newblock \emph{Astron. \& Astrophys.} 527, A70.
\newblock \doi{10.1051/0004-6361/201015777}
\bibAnnoteFile{2011A&A...527A..70K}

\bibitem[{{Kyutoku} and {Seto}(2016)}]{2016MNRAS.462.2177K}
{Kyutoku}, K. and {Seto}, N. (2016).
\newblock {Concise estimate of the expected number of detections for
  stellar-mass binary black holes by eLISA}.
\newblock \emph{Mon. Not. R. Ast. Soc.} 462, 2177--2183.
\newblock \doi{10.1093/mnras/stw1767}
\bibAnnoteFile{2016MNRAS.462.2177K}

\bibitem[{{Kyutoku} and {Seto}(2017)}]{2017PhRvD..95h3525K}
{Kyutoku}, K. and {Seto}, N. (2017).
\newblock {Gravitational-wave cosmography with LISA and the Hubble tension}.
\newblock \emph{Phys. Rev. D} 95, 083525.
\newblock \doi{10.1103/PhysRevD.95.083525}
\bibAnnoteFile{2017PhRvD..95h3525K}

\bibitem[{{Lau} et~al.(2020){Lau}, {Mandel}, {Vigna-G{\'o}mez}, {Neijssel},
  {Stevenson}, and {Sesana}}]{2020MNRAS.492.3061L}
{Lau}, M. Y.~M., {Mandel}, I., {Vigna-G{\'o}mez}, A., {Neijssel}, C.~J.,
  {Stevenson}, S., and {Sesana}, A. (2020).
\newblock {Detecting double neutron stars with LISA}.
\newblock \emph{Mon. Not. R. Astron. Soc.} 492, 3061--3072.
\newblock \doi{10.1093/mnras/staa002}
\bibAnnoteFile{2020MNRAS.492.3061L}

\bibitem[{{Levin}(2007)}]{2007MNRAS.374..515L}
{Levin}, Y. (2007).
\newblock {Starbursts near supermassive black holes: young stars in the
  Galactic Centre, and gravitational waves in LISA band}.
\newblock \emph{Mon. Not. R. Ast. Soc.} 374, 515--524.
\newblock \doi{10.1111/j.1365-2966.2006.11155.x}
\bibAnnoteFile{2007MNRAS.374..515L}

\bibitem[{{LSST Science Collaboration} et~al.(2009){LSST Science
  Collaboration}, {Abell}, {Allison}, {Anderson}, {Andrew}, {Angel}
  et~al.}]{2009arXiv0912.0201L}
{LSST Science Collaboration}, {Abell}, P.~A., {Allison}, J., {Anderson}, S.~F.,
  {Andrew}, J.~R., {Angel}, J. R.~P., et~al. (2009).
\newblock {LSST Science Book, Version 2.0}.
\newblock \emph{arXiv e-prints} , arXiv:0912.0201
\bibAnnoteFile{2009arXiv0912.0201L}

\bibitem[{{MacLeod} and {Hogan}(2008)}]{2008PhRvD..77d3512M}
{MacLeod}, C.~L. and {Hogan}, C.~J. (2008).
\newblock {Precision of Hubble constant derived using black hole binary
  absolute distances and statistical redshift information}.
\newblock \emph{Phys. Rev. D} 77, 043512.
\newblock \doi{10.1103/PhysRevD.77.043512}
\bibAnnoteFile{2008PhRvD..77d3512M}

\bibitem[{{Madau} and {Rees}(2001)}]{2001ApJ...551L..27M}
{Madau}, P. and {Rees}, M.~J. (2001).
\newblock {Massive Black Holes as Population III Remnants}.
\newblock \emph{Astrophys. J. Lett.} 551, L27--L30.
\newblock \doi{10.1086/319848}
\bibAnnoteFile{2001ApJ...551L..27M}

\bibitem[{{Mangiagli} et~al.(2020){Mangiagli}, {Klein}, {Bonetti}, {Katz},
  {Sesana}, {Volonteri} et~al.}]{2020arXiv200612513M}
{Mangiagli}, A., {Klein}, A., {Bonetti}, M., {Katz}, M.~L., {Sesana}, A.,
  {Volonteri}, M., et~al. (2020).
\newblock {On the inspiral of coalescing massive black hole binaries with LISA
  in the era of Multi-Messenger Astrophysics}.
\newblock \emph{arXiv e-prints} , arXiv:2006.12513
\bibAnnoteFile{2020arXiv200612513M}

\bibitem[{{McGee} et~al.(2020){McGee}, {Sesana}, and
  {Vecchio}}]{2020NatAs...4...26M}
{McGee}, S., {Sesana}, A., and {Vecchio}, A. (2020).
\newblock {Linking gravitational waves and X-ray phenomena with joint LISA and
  Athena observations}.
\newblock \emph{Nature Astronomy} 4, 26--31.
\newblock \doi{10.1038/s41550-019-0969-7}
\bibAnnoteFile{2020NatAs...4...26M}

\bibitem[{{McWilliams} et~al.(2010){McWilliams}, {Thorpe}, {Baker}, and
  {Kelly}}]{2010PhRvD..81f4014M}
{McWilliams}, S.~T., {Thorpe}, J.~I., {Baker}, J.~G., and {Kelly}, B.~J.
  (2010).
\newblock {Impact of mergers on LISA parameter estimation for nonspinning black
  hole binaries}.
\newblock \emph{Phys. Rev. D} 81, 064014.
\newblock \doi{10.1103/PhysRevD.81.064014}
\bibAnnoteFile{2010PhRvD..81f4014M}

\bibitem[{{Milosavljevi{\'c}} and {Phinney}(2005)}]{2005ApJ...622L..93M}
{Milosavljevi{\'c}}, M. and {Phinney}, E.~S. (2005).
\newblock {The Afterglow of Massive Black Hole Coalescence}.
\newblock \emph{Astrophys. J. Lett.} 622, L93--L96.
\newblock \doi{10.1086/429618}
\bibAnnoteFile{2005ApJ...622L..93M}

\bibitem[{{Miyoshi} et~al.(1995){Miyoshi}, {Moran}, {Herrnstein}, {Greenhill},
  {Nakai}, {Diamond} et~al.}]{1995Natur.373..127M}
{Miyoshi}, M., {Moran}, J., {Herrnstein}, J., {Greenhill}, L., {Nakai}, N.,
  {Diamond}, P., et~al. (1995).
\newblock {Evidence for a black hole from high rotation velocities in a
  sub-parsec region of NGC4258}.
\newblock \emph{Nature} 373, 127--129.
\newblock \doi{10.1038/373127a0}
\bibAnnoteFile{1995Natur.373..127M}

\bibitem[{{Nelemans} et~al.(2001){Nelemans}, {Yungelson}, and {Portegies
  Zwart}}]{2001A&A...375..890N}
{Nelemans}, G., {Yungelson}, L.~R., and {Portegies Zwart}, S.~F. (2001).
\newblock {The gravitational wave signal from the Galactic disk population of
  binaries containing two compact objects}.
\newblock \emph{Astron. \& Astrophys.} 375, 890--898.
\newblock \doi{10.1051/0004-6361:20010683}
\bibAnnoteFile{2001A&A...375..890N}

\bibitem[{{Nishizawa} et~al.(2016){Nishizawa}, {Berti}, {Klein}, and
  {Sesana}}]{2016PhRvD..94f4020N}
{Nishizawa}, A., {Berti}, E., {Klein}, A., and {Sesana}, A. (2016).
\newblock {eLISA eccentricity measurements as tracers of binary black hole
  formation}.
\newblock \emph{Phys. Rev. D} 94, 064020.
\newblock \doi{10.1103/PhysRevD.94.064020}
\bibAnnoteFile{2016PhRvD..94f4020N}

\bibitem[{{Nishizawa} et~al.(2017){Nishizawa}, {Sesana}, {Berti}, and
  {Klein}}]{2017MNRAS.465.4375N}
{Nishizawa}, A., {Sesana}, A., {Berti}, E., and {Klein}, A. (2017).
\newblock {Constraining stellar binary black hole formation scenarios with
  eLISA eccentricity measurements}.
\newblock \emph{Mon. Not. R. Ast. Soc.} 465, 4375--4380.
\newblock \doi{10.1093/mnras/stw2993}
\bibAnnoteFile{2017MNRAS.465.4375N}

\bibitem[{{Nissanke} et~al.(2012){Nissanke}, {Vallisneri}, {Nelemans}, and
  {Prince}}]{2012ApJ...758..131N}
{Nissanke}, S., {Vallisneri}, M., {Nelemans}, G., and {Prince}, T.~A. (2012).
\newblock {Gravitational-wave Emission from Compact Galactic Binaries}.
\newblock \emph{Astrophys. J.} 758, 131.
\newblock \doi{10.1088/0004-637X/758/2/131}
\bibAnnoteFile{2012ApJ...758..131N}

\bibitem[{{Palenzuela} et~al.(2010){Palenzuela}, {Lehner}, and
  {Liebling}}]{2010Sci...329..927P}
{Palenzuela}, C., {Lehner}, L., and {Liebling}, S.~L. (2010).
\newblock {Dual Jets from Binary Black Holes}.
\newblock \emph{Science} 329, 927--930.
\newblock \doi{10.1126/science.1191766}
\bibAnnoteFile{2010Sci...329..927P}

\bibitem[{{Perego} et~al.(2009){Perego}, {Dotti}, {Colpi}, and
  {Volonteri}}]{2009MNRAS.399.2249P}
{Perego}, A., {Dotti}, M., {Colpi}, M., and {Volonteri}, M. (2009).
\newblock {Mass and spin co-evolution during the alignment of a black hole in a
  warped accretion disc}.
\newblock \emph{Mon. Not. R. Ast. Soc.} 399, 2249--2263.
\newblock \doi{10.1111/j.1365-2966.2009.15427.x}
\bibAnnoteFile{2009MNRAS.399.2249P}

\bibitem[{{Punturo} et~al.(2010){Punturo}, {Abernathy}, {Acernese}, {Allen},
  {Andersson}, {Arun} et~al.}]{2010CQGra..27s4002P}
{Punturo}, M., {Abernathy}, M., {Acernese}, F., {Allen}, B., {Andersson}, N.,
  {Arun}, K., et~al. (2010).
\newblock {The Einstein Telescope: a third-generation gravitational wave
  observatory}.
\newblock \emph{Classical and Quantum Gravity} 27, 194002.
\newblock \doi{10.1088/0264-9381/27/19/194002}
\bibAnnoteFile{2010CQGra..27s4002P}

\bibitem[{{Ravi} et~al.(2012){Ravi}, {Wyithe}, {Hobbs}, {Shannon},
  {Manchester}, {Yardley} et~al.}]{2012ApJ...761...84R}
{Ravi}, V., {Wyithe}, J.~S.~B., {Hobbs}, G., {Shannon}, R.~M., {Manchester},
  R.~N., {Yardley}, D.~R.~B., et~al. (2012).
\newblock {Does a ``Stochastic'' Background of Gravitational Waves Exist in the
  Pulsar Timing Band?}
\newblock \emph{Astrophys. J.} 761, 84.
\newblock \doi{10.1088/0004-637X/761/2/84}
\bibAnnoteFile{2012ApJ...761...84R}

\bibitem[{{Reitze} et~al.(2019){Reitze}, {Adhikari}, {Ballmer}, {Barish},
  {Barsotti}, {Billingsley} et~al.}]{2019BAAS...51g..35R}
{Reitze}, D., {Adhikari}, R.~X., {Ballmer}, S., {Barish}, B., {Barsotti}, L.,
  {Billingsley}, G., et~al. (2019).
\newblock {Cosmic Explorer: The U.S. Contribution to Gravitational-Wave
  Astronomy beyond LIGO}.
\newblock In \emph{Bulletin of the American Astronomical Society}. vol.~51, 35
\bibAnnoteFile{2019BAAS...51g..35R}

\bibitem[{{Reynolds}(2014)}]{2014SSRv..183..277R}
{Reynolds}, C.~S. (2014).
\newblock {Measuring Black Hole Spin Using X-Ray Reflection Spectroscopy}.
\newblock \emph{Space Sci. Rev.} 183, 277--294.
\newblock \doi{10.1007/s11214-013-0006-6}
\bibAnnoteFile{2014SSRv..183..277R}

\bibitem[{{Roedig} et~al.(2014){Roedig}, {Krolik}, and
  {Miller}}]{2014ApJ...785..115R}
{Roedig}, C., {Krolik}, J.~H., and {Miller}, M.~C. (2014).
\newblock {Observational Signatures of Binary Supermassive Black Holes}.
\newblock \emph{Astrophys. J.} 785, 115.
\newblock \doi{10.1088/0004-637X/785/2/115}
\bibAnnoteFile{2014ApJ...785..115R}

\bibitem[{{Rossi} et~al.(2010){Rossi}, {Lodato}, {Armitage}, {Pringle}, and
  {King}}]{2010MNRAS.401.2021R}
{Rossi}, E.~M., {Lodato}, G., {Armitage}, P.~J., {Pringle}, J.~E., and {King},
  A.~R. (2010).
\newblock {Black hole mergers: the first light}.
\newblock \emph{Mon. Not. R. Ast. Soc.} 401, 2021--2035.
\newblock \doi{10.1111/j.1365-2966.2009.15802.x}
\bibAnnoteFile{2010MNRAS.401.2021R}

\bibitem[{{Samsing} and {D'Orazio}(2018)}]{2018MNRAS.481.5445S}
{Samsing}, J. and {D'Orazio}, D.~J. (2018).
\newblock {Black Hole Mergers From Globular Clusters Observable by LISA I:
  Eccentric Sources Originating From Relativistic N-body Dynamics}.
\newblock \emph{Mon. Not. R. Ast. Soc.} 481, 5445--5450.
\newblock \doi{10.1093/mnras/sty2334}
\bibAnnoteFile{2018MNRAS.481.5445S}

\bibitem[{{Sesana}(2013)}]{2013ASPC..467..103S}
{Sesana}, A. (2013).
\newblock {Detecting Massive Black Hole Binaries and Unveiling their Cosmic
  History with Gravitational Wave Observations}.
\newblock In \emph{9th LISA Symposium}, eds. G.~{Auger}, P.~{Bin{\'e}truy}, and
  E.~{Plagnol}. vol. 467 of \emph{Astronomical Society of the Pacific
  Conference Series}, 103
\bibAnnoteFile{2013ASPC..467..103S}

\bibitem[{{Sesana}(2016)}]{2016PhRvL.116w1102S}
{Sesana}, A. (2016).
\newblock {Prospects for Multiband Gravitational-Wave Astronomy after
  GW150914}.
\newblock \emph{Physical Review Letters} 116, 231102.
\newblock \doi{10.1103/PhysRevLett.116.231102}
\bibAnnoteFile{2016PhRvL.116w1102S}

\bibitem[{{Sesana}(2017)}]{2017JPhCS.840a2018S}
{Sesana}, A. (2017).
\newblock {Multi-band gravitational wave astronomy: science with joint space-
  and ground-based observations of black hole binaries}.
\newblock In \emph{Journal of Physics Conference Series}. vol. 840 of
  \emph{Journal of Physics Conference Series}, 012018.
\newblock \doi{10.1088/1742-6596/840/1/012018}
\bibAnnoteFile{2017JPhCS.840a2018S}

\bibitem[{{Sesana} et~al.(2011){Sesana}, {Gair}, {Berti}, and
  {Volonteri}}]{2011PhRvD..83d4036S}
{Sesana}, A., {Gair}, J., {Berti}, E., and {Volonteri}, M. (2011).
\newblock {Reconstructing the massive black hole cosmic history through
  gravitational waves}.
\newblock \emph{Phys. Rev. D} 83, 044036.
\newblock \doi{10.1103/PhysRevD.83.044036}
\bibAnnoteFile{2011PhRvD..83d4036S}

\bibitem[{{Sesana} et~al.(2020){Sesana}, {Lamberts}, and
  {Petiteau}}]{2020MNRAS.494L..75S}
{Sesana}, A., {Lamberts}, A., and {Petiteau}, A. (2020).
\newblock {Finding binary black holes in the Milky Way with LISA}.
\newblock \emph{Mon. Not. R. Ast. Soc.} 494, L75--L80.
\newblock \doi{10.1093/mnrasl/slaa039}
\bibAnnoteFile{2020MNRAS.494L..75S}

\bibitem[{{Sesana} et~al.(2012){Sesana}, {Roedig}, {Reynolds}, and
  {Dotti}}]{2012MNRAS.420..860S}
{Sesana}, A., {Roedig}, C., {Reynolds}, M.~T., and {Dotti}, M. (2012).
\newblock {Multimessenger astronomy with pulsar timing and X-ray observations
  of massive black hole binaries}.
\newblock \emph{Mon. Not. R. Ast. Soc.} 420, 860--877.
\newblock \doi{10.1111/j.1365-2966.2011.20097.x}
\bibAnnoteFile{2012MNRAS.420..860S}

\bibitem[{{Sesana} et~al.(2008){Sesana}, {Vecchio}, and
  {Colacino}}]{2008MNRAS.390..192S}
{Sesana}, A., {Vecchio}, A., and {Colacino}, C.~N. (2008).
\newblock {The stochastic gravitational-wave background from massive black hole
  binary systems: implications for observations with Pulsar Timing Arrays}.
\newblock \emph{Mon. Not. R. Ast. Soc.} 390, 192--209.
\newblock \doi{10.1111/j.1365-2966.2008.13682.x}
\bibAnnoteFile{2008MNRAS.390..192S}

\bibitem[{{Sesana} et~al.(2009){Sesana}, {Vecchio}, and
  {Volonteri}}]{2009MNRAS.394.2255S}
{Sesana}, A., {Vecchio}, A., and {Volonteri}, M. (2009).
\newblock {Gravitational waves from resolvable massive black hole binary
  systems and observations with Pulsar Timing Arrays}.
\newblock \emph{Mon. Not. R. Ast. Soc.} 394, 2255--2265.
\newblock \doi{10.1111/j.1365-2966.2009.14499.x}
\bibAnnoteFile{2009MNRAS.394.2255S}

\bibitem[{{Seto}(2016)}]{2016MNRAS.460L...1S}
{Seto}, N. (2016).
\newblock {Prospects of eLISA for detecting Galactic binary black holes similar
  to GW150914}.
\newblock \emph{Mon. Not. R. Ast. Soc.} 460, L1--L4.
\newblock \doi{10.1093/mnrasl/slw060}
\bibAnnoteFile{2016MNRAS.460L...1S}

\bibitem[{{Shapiro} et~al.(2010){Shapiro}, {Bacon}, {Hendry}, and
  {Hoyle}}]{2010MNRAS.404..858S}
{Shapiro}, C., {Bacon}, D.~J., {Hendry}, M., and {Hoyle}, B. (2010).
\newblock {Delensing gravitational wave standard sirens with shear and flexion
  maps}.
\newblock \emph{Mon. Not. R. Ast. Soc.} 404, 858--866.
\newblock \doi{10.1111/j.1365-2966.2010.16317.x}
\bibAnnoteFile{2010MNRAS.404..858S}

\bibitem[{{Tamanini} et~al.(2016){Tamanini}, {Caprini}, {Barausse}, {Sesana},
  {Klein}, and {Petiteau}}]{2016JCAP...04..002T}
{Tamanini}, N., {Caprini}, C., {Barausse}, E., {Sesana}, A., {Klein}, A., and
  {Petiteau}, A. (2016).
\newblock {Science with the space-based interferometer eLISA. III: probing the
  expansion of the universe using gravitational wave standard sirens}.
\newblock \emph{J. of Cosm. and Astropart. Phys.} 4, 002.
\newblock \doi{10.1088/1475-7516/2016/04/002}
\bibAnnoteFile{2016JCAP...04..002T}

\bibitem[{{Tamanini} and {Danielski}(2019)}]{2019NatAs...3..858T}
{Tamanini}, N. and {Danielski}, C. (2019).
\newblock {The gravitational-wave detection of exoplanets orbiting white dwarf
  binaries using LISA}.
\newblock \emph{Nature Astronomy} 3, 858--866.
\newblock \doi{10.1038/s41550-019-0807-y}
\bibAnnoteFile{2019NatAs...3..858T}

\bibitem[{{Tanaka} et~al.(2012){Tanaka}, {Menou}, and
  {Haiman}}]{2012MNRAS.420..705T}
{Tanaka}, T., {Menou}, K., and {Haiman}, Z. (2012).
\newblock {Electromagnetic counterparts of supermassive black hole binaries
  resolved by pulsar timing arrays}.
\newblock \emph{Mon. Not. R. Ast. Soc.} 420, 705--719.
\newblock \doi{10.1111/j.1365-2966.2011.20083.x}
\bibAnnoteFile{2012MNRAS.420..705T}

\bibitem[{{Tang} et~al.(2018){Tang}, {Haiman}, and
  {MacFadyen}}]{2018MNRAS.476.2249T}
{Tang}, Y., {Haiman}, Z., and {MacFadyen}, A. (2018).
\newblock {The late inspiral of supermassive black hole binaries with
  circumbinary gas discs in the LISA band}.
\newblock \emph{Mon. Not. R. Ast. Soc.} 476, 2249--2257.
\newblock \doi{10.1093/mnras/sty423}
\bibAnnoteFile{2018MNRAS.476.2249T}

\bibitem[{{Thorne}(1974)}]{1974ApJ...191..507T}
{Thorne}, K.~S. (1974).
\newblock {Disk-Accretion onto a Black Hole. II. Evolution of the Hole}.
\newblock \emph{Astrophys. J.} 191, 507--520.
\newblock \doi{10.1086/152991}
\bibAnnoteFile{1974ApJ...191..507T}

\bibitem[{{Tso} et~al.(2018){Tso}, {Gerosa}, and {Chen}}]{2018arXiv180700075T}
{Tso}, R., {Gerosa}, D., and {Chen}, Y. (2018).
\newblock {Optimizing LIGO with LISA forewarnings to improve black-hole
  spectroscopy}.
\newblock \emph{arXiv e-prints}
\bibAnnoteFile{2018arXiv180700075T}

\bibitem[{{Van Den Broeck}(2014)}]{2014JPhCS.484a2008V}
{Van Den Broeck}, C. (2014).
\newblock {Astrophysics, cosmology, and fundamental physics with compact binary
  coalescence and the Einstein Telescope}.
\newblock In \emph{Journal of Physics Conference Series}. vol. 484 of
  \emph{Journal of Physics Conference Series}, 012008.
\newblock \doi{10.1088/1742-6596/484/1/012008}
\bibAnnoteFile{2014JPhCS.484a2008V}

\bibitem[{{Verbiest} et~al.(2016){Verbiest}, {Lentati}, {Hobbs}, {van
  Haasteren}, {Demorest}, {Janssen} et~al.}]{2016MNRAS.458.1267V}
{Verbiest}, J.~P.~W., {Lentati}, L., {Hobbs}, G., {van Haasteren}, R.,
  {Demorest}, P.~B., {Janssen}, G.~H., et~al. (2016).
\newblock {The International Pulsar Timing Array: First data release}.
\newblock \emph{Mon. Not. R. Ast. Soc.} 458, 1267--1288.
\newblock \doi{10.1093/mnras/stw347}
\bibAnnoteFile{2016MNRAS.458.1267V}

\bibitem[{{Volonteri} et~al.(2003){Volonteri}, {Haardt}, and
  {Madau}}]{2003ApJ...582..559V}
{Volonteri}, M., {Haardt}, F., and {Madau}, P. (2003).
\newblock {The Assembly and Merging History of Supermassive Black Holes in
  Hierarchical Models of Galaxy Formation}.
\newblock \emph{Astrophys. J.} 582, 559--573.
\newblock \doi{10.1086/344675}
\bibAnnoteFile{2003ApJ...582..559V}

\bibitem[{{Woods} et~al.(2018){Woods}, {Agarwal}, {Bromm}, {Bunker}, {Chen},
  {Chon} et~al.}]{2018arXiv181012310W}
{Woods}, T.~E., {Agarwal}, B., {Bromm}, V., {Bunker}, A., {Chen}, K.-J.,
  {Chon}, S., et~al. (2018).
\newblock {Titans of the Early Universe: The Prato Statement on the Origin of
  the First Supermassive Black Holes}.
\newblock \emph{arXiv e-prints}
\bibAnnoteFile{2018arXiv181012310W}

\end{thebibliography}

\end{document}